%% file: main.tex
\documentclass[prl,aps,twocolumn,preprintnumbers,amssymb,nobibnotes,nofootinbib,raggedbottom,10pt]{revtex4-2}
\usepackage{amsmath}
\usepackage{hyperref,graphicx,color,amsmath,dsfont,titlesec,shuffle,tikz}
\usepackage{upgreek}
\usepackage{mathtools}
\usepackage[most]{tcolorbox}
\input{header_data.tex}

\newcommand{\Ddots}{\hbox to 1em{.\hss.\hss.\hss}}

\begin{document}
\title {Universality of Colored Scalars from the Stringy KLT Kernel}
\author{Christoph Bartsch$^1$}
\author{Karol Kampf$\,^{1}$}
\author{Ji\v r\' i Novotn\' y$^1$}
\author{Jaroslav Trnka$^{2}$}

\affiliation{$^1$Institute for Particle and Nuclear Physics, Charles University, Prague, Czech Republic}
\affiliation{$^2$Center for Quantum Mathematics and Physics (QMAP), University of California, Davis, CA, USA}

\begin{abstract}
    \noindent A new perspective on the inverse string theory Kawai-Lewellen-Tye (KLT) kernel is provided which establishes the universality of scattering amplitudes in the bi-adjoint scalar (BAS) theory, pions in the Non-linear sigma model (NLSM), and mixed amplitudes (NLSM+$\phi^3$) recently studied in the literature. We show that all these amplitudes can be viewed as equivalent, arising from a single function, the inverse string theory KLT kernel, evaluated at different kinematic points. In this way cubic colored scalars and pions become interchangeable through a procedure we call the $\alpha'$-shift. The latter complements the $\delta$-shift proposed by Arkani-Hamed et al., and demonstrates an inherent equivalence of scattering amplitudes in different quantum field theories by embedding them in a common stringy framework.
\end{abstract}

\maketitle

\vspace{-0.3cm}

\subparagraph{I. Introduction}

String-inspired techniques play an important role in discovering new properties of scattering amplitudes for point particles. In particular, a lot of progress was made recently in understanding close relationships between amplitudes of bi-adjoint scalars (BAS), pions in the non-linear sigma model (NLSM) and gluons in Yang-Mills theory from the kinematic surface picture \cite{Arkani-Hamed:2024pzc,Arkani-Hamed:2023jry,Arkani-Hamed:2024tzl}.
At the heart of their unity lies a kinematic deformation, the $\delta$-shift, which parametrizes a family of worldsheet integrals yielding the desired amplitudes in various limits \cite{Arkani-Hamed:2024vna,Arkani-Hamed:2023swr}.
The connection is particularly immediate for scalars as field theory amplitudes for pions can be directly extracted from those of cubic scalars in the BAS theory, even at loop-level \cite{Arkani-Hamed:2024yvu,Arkani-Hamed:2024nhp}.

Another well-studied connection between BAS and NLSM amplitudes is via their stringy extensions into (non-)abelian $Z$-functions \cite{Broedel:2013tta,Mafra:2016mcc,Carrasco:2016ygv,Carrasco:2016ldy}. There, relations between cubic scalars and pions, as well as mixed amplitudes involving both (NLSM+BAS) are established via a procedure known as (semi-)abelianization.

All the aforementioned amplitudes share another string-derived property: the duality between color and kinematics. As a consequence, amplitudes of cubic scalars, pions and gluons, among others, can be used as input in a procedure known as the double copy to generate permutationally invariant amplitudes for gravity, the special Galileon and more \cite{Bern:2019prr}. A central object in this context is the string theory KLT kernel \cite{Bjerrum-Bohr:2010pnr} $\mathcal{S}_{\alpha'}\!\left[\sigma | \rho \right]$
which was first discovered by observing the double-copy structure of closed and open string amplitudes \cite{Kawai:1985xq},
\begin{align*}
    M^{\text{closed}} =\!\!\!\! \sum_{\sigma,\rho \in S_{n-3}}\!\!\!\! A^{\text{open}}\left[\sigma\right] \mathcal{S}_{\alpha'}\!\left[\sigma | \rho \right] A^{\text{open}}\left[\rho\right].
\end{align*}
In this letter we set out to connect the world of the $\delta$-shift and that of the (semi-)abelianization procedures established in the context of (non-)abelian $Z$-theory. This will be accomplished by studying the simplest known stringy extension of BAS amplitudes provided by the inverse $m^{\alpha'}\equiv \mathcal{S}_{\alpha'}^{-1} $ of the string theory KLT kernel \cite{Mizera:2016jhj,Mizera:2017cqs}.
Consequently, we will establish that the KLT kernel encodes not only the scattering of cubic scalars but also pion and mixed amplitudes in the NLSM+BAS theory.

To extract this information we will first study their connection by analogy to $Z$-theory using (semi-) abelianization to obtain stringy functions related to the scattering of pions. A simple general structure observed in these functions then naturally suggests the construction of a stringy variant of the $\delta$-shift, which we call the $\alpha'$-shift. We will argue that the $\alpha'$-shift establishes a fundamental equivalence between cubic scalars and pions which may be most succinctly summarized as
\begin{equation*}
\text{BAS}_{\alpha'} = \text{NLSM}_{\alpha'} = (\text{NLSM}+\text{BAS})_{\alpha'} = \text{KLT}_{\alpha'}^{-1}.
\end{equation*}
That is to say, all amplitudes for scalars and pions can be obtained from the \textit{same function}, the inverse string theory KLT kernel, evaluated on appropriately $\alpha'$-shifted kinematics.

\subparagraph{II. The Inverse String Theory KLT Kernel}

In \cite{Mizera:2016jhj,Mizera:2017cqs} Mizera initiated the study of the matrix inverse of the KLT kernel $m^{\alpha'} \!\equiv \mathcal{S}_{\alpha'}^{-1}$ and showed that its matrix elements $m^{\alpha'}\!\left[ \sigma | \rho \right]$ are simple kinematic functions. Explicit examples for these matrix elements include the three-point function $m_3^{\alpha'}\left[\mathds{1}|\mathds{1}\right] = 1$ as well as
\begin{align}
\begin{split}\label{invKLTex}
     &\hspace{-0.20cm}m_4^{\alpha'}\!\!\left[\mathds{1}|\mathds{1}\right] = \frac{1}{t_{13}}{+}\frac{1}{t_{24}}, \hspace{0.1cm} m_5^{\alpha'}\!\!\left[\mathds{1}|\mathds{1}\right] = \left(\!\frac{1}{t_{13}t_{14}} {+} \text{cyc.}\!\right) {+}1,
    \\
    &\hspace{-0.20cm}m_4^{\alpha'}\!\!\left[\mathds{1}|1243\right] = -\frac{1}{s_{13}}, \hspace{0.1cm}
    m_5^{\alpha'}\!\!\left[\mathds{1}|13245\right] =\frac{1}{s_{14}}\!\left(\!\frac{1}{t_{13}}{+}\frac{1}{t_{24}}\!\right)\!,
\end{split}
\end{align}
at four and five points. We denote by $\mathds{1}$ the identity permutation of $n$ labels and introduce the $\alpha'$-dependent planar Mandelstam variables
\begin{align}\label{strPlMand}
t_{ij}=\tan\left(\pi\alpha' X_{ij}\right), \hspace{0.5cm} s_{ij}=\sin\left(\pi\alpha' X_{ij} \right),
\end{align}
which are stringy versions of the field theory $X$-variables $X_{ij}=(p_i+\dots+p_{j-1})^2$.
From \eqref{invKLTex} we see that diagonal matrix elements $m_n^{\alpha'}\!\left[\mathds{1}|\mathds{1}\right]$ depend only on $t_{ij}$ whereas off-diagonal elements $m_n^{\alpha'}\!\left[\mathds{1}|\rho\right]$, $\rho\neq\mathds{1}$ generically involve $t_{ij}$ and $s_{ij}$.

Taking the infinite tension limit $\alpha'{\to}\, 0$ of $m_n^{\alpha'}$ yields partial amplitudes of the bi-adjoint scalar (BAS) theory,
\begin{align} \label{alpha0limBAS}
    m_n^{\alpha'}\!\left[\sigma | \rho \right] = (\pi\alpha')^{3-n}\big(m_n\left[\sigma|\rho\right] + \mathcal{O}(\alpha')\big).
\end{align}
A crucial feature of the stringy matrix elements $m_n^{\alpha'}$ is the presence of an infinite tower of odd-point interactions \cite{Mizera:2016jhj} (c.f. the five-point contact term in \eqref{invKLTex}). In fact, diagonal matrix elements $m_n^{\alpha'}\!\left[\mathds{1}|\mathds{1}\right]$ can be computed as sums over all Feynman diagrams constructed from the given odd-point vertices and propagators $t_{ij}^{-1}$.
For instance, at six points, the diagonal matrix element is schematically given by the diagrams
\vspace{-0.1cm}
\begin{align}\label{6ptDiag}
    \begin{matrix}
        \hspace{-0.28cm}\vspace{0.0cm}\includegraphics[width=8.10cm]{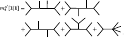}
        \end{matrix}
\end{align}

\vspace{-0.1cm}
In the limit $\alpha'\to 0$ only the cubic topologies contribute to the field theory BAS amplitude $m_6\left[\mathds{1}|\mathds{1}\right]$ by \eqref{alpha0limBAS}, while the topology involving the five-point vertex is sub-leading. Nevertheless, sub-leading-$\alpha'$ corrections in the inverse string theory KLT kernel $m_n^{\alpha'}$ will be essential to connect to pions and mixed amplitudes.

Given the diagram topologies for $m_n\left[\mathds{1}|\mathds{1}\right]$, off-diagonal matrix elements $m_n^{\alpha'}\!\left[\mathds{1}|\rho\right]$ then correspond to subsets of those diagrams which are planar with respect to both orderings $\mathds{1}$ and $\rho$, analogous to off-diagonal BAS amplitudes \cite{Cachazo:2013iea}. In addition, certain stringy propagators $t_{ij}^{-1}$ are replaced by $s_{ij}^{-1}$ for the off-diagonal matrix elements.

\subparagraph{III. Pions via Abelianization}
Let us now define an \textit{abelianized} function
\begin{align} \label{defAbel}
    A_n^{\alpha'}\!\left[\sigma \right] \equiv \frac{1}{2}\!\sum_{\rho(2\dots n)}\! \!m_n^{\alpha'}\!\left[\sigma|1\,\rho\right],
\end{align}
as a sum over one of the orderings of the inverse KLT kernel. Here we adopt the nomenclature of the $Z$-theory literature \cite{Broedel:2013tta,Mafra:2016mcc,Carrasco:2016ygv,Carrasco:2016ldy}, where the notion of abelianization relates so-called non-abelian and abelian $Z$-functions. Both are closely related to the inverse KLT kernel \cite{Mizera:2016jhj}.

A simple example of \eqref{defAbel} is provided by the abelianized four-point function,
\begin{align}\label{abel4pt}
    A^{\alpha'}_4\!\left[\mathds{1}\right] =\! \left(\!\frac{1}{t_{13}}{-}\frac{1}{s_{13}}\!\right){+}\left(\!\frac{1}{t_{24}}{-}\frac{1}{s_{24}}\!\right) = -\left(\tau_{13}+\tau_{24}\right).
\end{align}
Here we introduce the half-angle variables
\begin{align}
\tau_{ij}=\tan\left(\frac{\pi}{2}\alpha' X_{ij}\right),
\end{align}
which are the natural kinematic variables for abelianized functions, as we will see. They are related to the variables $t_{ij}, \,s_{ij}$ of the inverse KLT kernel via trigonometric half-angle identities
\begin{align}\label{trigID1}
    \frac{1}{t_{ij}} + \frac{1}{s_{ij}}= \frac{1}{\tau_{ij}}, \hspace{0.5cm} \frac{1}{t_{ij}} - \frac{1}{s_{ij}}= -\tau_{ij}.
\end{align}
Expanding the four-point function \eqref{abel4pt} in powers of $\alpha'$ we obtain $A^{\alpha'}_4 = -\frac{\pi\alpha'}{2}(X_{13}+X_{24}) + \mathcal{O}(\alpha'^3)$ at leading order which is the four-pion NLSM amplitude.

This holds true more generally. For even multiplicity $n=2k$ the low-energy limit of the abelianized functions \eqref{defAbel} is given by NLSM pion amplitudes,
\begin{align}\label{alpha0limNLSM}
    A_{2k}^{\alpha'} = \frac{\pi\alpha'}{2}A_{2k}^{\text{NLSM}} + \mathcal{O}(\alpha'^3).
\end{align}

An analogous statement to \eqref{alpha0limNLSM} also holds for the leading-$\alpha'$ contribution to abelian $Z$-functions \cite{Carrasco:2016ldy}. However, the abelianized functions $A_{2k}^{\alpha'}$ provide a stringy completion of pion amplitudes that is distinct from abelian $Z$-theory at higher orders in $\alpha'$. In particular, $Z$-functions are defined in terms of disk integrals for which analytic results in $\alpha'$ are hard to obtain when $n\ge5$.

On the other hand, the abelian functions \eqref{defAbel} turn out to be simple rational functions of the stringy variables $\tau_{ij}$. For $n=2k$ even, their structure is closely related to that of field theory pion amplitudes in the NLSM. Considering pion amplitudes $A_{2k}^{\text{NLSM}}(X_{ij})$ directly as functions of invariants $X_{ij}$ the general functional form of the abelianized function \eqref{defAbel} is schematically given by
\begin{align}\label{abelFromNLSM}
    A^{\alpha'}_{2k}(\tau_{ij}) = A_{2k}^{\text{NLSM}}(\tau_{ij}) + \text{higher order in }\tau_{ij}.
\end{align}
That is, the leading $\mathcal{O}(\tau_{ij})$ contribution to the stringy pion functions \eqref{abelFromNLSM} can be obtained by taking the field theory amplitude $A_{2k}^{\text{NLSM}}(X_{ij})$ and replacing $X_{ij}\mapsto \tau_{ij}$.
For example, the four-point function \eqref{abel4pt} can be seen as a realization of the general structure \eqref{abelFromNLSM} where higher-order corrections in $\tau_{ij}$ are absent.

More broadly, the stringy pion functions \eqref{abelFromNLSM} exhibit various properties familiar from their field theory analogues in the NLSM. For one, they consistently factorize as functions of $\tau_{ij}$ on poles corresponding to odd-particle channels,
\begin{align}
    \mathop{\mathrm{Res}}_{\tau_{\text{odd}}=0} A_{2k}^{\alpha'}\!(\tau) = A_L^{\alpha'}\!(\tau_L)\,A_R^{\alpha'}\!(\tau_R)
\end{align}
whereas there are no poles for even-particle channels at $\tau_{\text{even}}=0$.
Moreover, they obey the Adler zero \cite{Adler:1964um},
\begin{align}
    \lim_{p_i \to 0} A_{2k}^{\alpha'} = 0,
\end{align}
retaining a highly characteristic feature of pion amplitudes at finite $\alpha'$.

\subparagraph{Diagram-level abelianization: 6-point example}
We would now like to understand in more detail how pions arise from the abelianization of the inverse KLT kernel. Since the matrix elements $m_n^{\alpha'}$ are obtained from Feynman diagrams we want to study how the abelianization \eqref{defAbel} acts on individual diagram topologies. To illustrate, we consider an example at $n=6$ provided by the first diagram in \eqref{6ptDiag}. Its contribution to a matrix element $m_n^{\alpha'}\!\left[\mathds{1}|1\rho \right]$ on the right-hand side of \eqref{defAbel} is given by
\begin{align}\label{abel61}
        \begin{matrix}
        \vspace{-0.1cm}\includegraphics[width=3.8cm]{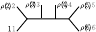}
        \end{matrix}\!\!\! = \frac{\theta(\rho)}{x_{13}(\rho)x_{14}(\rho)x_{15}(\rho)}\,.
\end{align}
Here $x_{ij}(\rho)$ equals $t_{ij}$ or $s_{ij}$ depending on $\rho$. As indicated before, a given diagram may not be planar with respect to the ordering $\rho$, in which case $\theta(\rho)=0$ and the diagram does not contribute to $m_n^{\alpha'}\!\left[\mathds{1}|1\rho \right]$. Otherwise, 
$\theta(\rho)$ 
is an overall sign depending on the specifics of $\rho$. In this case, we get a non-vanishing contribution to the matrix element $m_n^{\alpha'}\!\left[\mathds{1}|1\rho \right]$ from \eqref{abel61}.

Abelianizing the diagram amounts to summing \eqref{abel61} over permutations $\rho(2\dots n)$ as in \eqref{defAbel}. Although we naively expect $5!$ contributions, only a small number are not disappearing due to $\theta(\rho)\neq 0$.
Carrying out the sum we find a simple factorized form
\begin{align}\label{abel61res}
    \frac{1}{2}\!\!\!\sum_{\rho(2\dots 6)} \!\!\!\!\eqref{abel61} =\! \left(\!\frac{1}{t_{13}}{-}\frac{1}{s_{13}}\!\right)\!\!\!\left(\!\frac{1}{t_{14}}{+}\frac{1}{s_{14}}\!\right)\!\!\!\left(\!\frac{1}{t_{15}}{-}\frac{1}{s_{15}}\!\right) \!= \frac{\tau_{13}\tau_{15}}{\tau_{14}},
\end{align}
which can be expressed, like \eqref{abel4pt}, purely in terms of stringy variables $\tau_{ij}$ by using \eqref{trigID1}.

Proceeding to abelianize all remaining diagrams in \eqref{6ptDiag} we find a $\tau_{ij}$-form analogous to \eqref{abel61res} for each and combine them into the abelianized six-point function
\begin{align}\label{abel6pt}
    A_6^{\alpha'} \!\!= \frac{1}{2}\frac{(\tau_{13}+\tau_{24})(\tau_{46}+\tau_{15})}{\tau_{14}}-\tau_{13}-\frac{1}{3}\tau_{13}\tau_{35}\tau_{15} + \text{cyc.}
\end{align}

Here we recognize the abelianized diagram \eqref{abel61res} as contributing to the first term.
This shows that the abelianization is a sensible operation locally for each diagram, not just at the level of the full matrix element as in \eqref{defAbel}. Indeed, each term in \eqref{abel6pt} uniquely corresponds to a diagram in the diagonal inverse KLT kernel $m_6^{\alpha'}\!\left[\mathds{1}|\mathds{1}\right]$ \eqref{6ptDiag}.

In \eqref{abel6pt} we also easily recognize the general structure of \eqref{abelFromNLSM} once we compare it to the field theory NLSM amplitude \cite{Kampf:2013vha},
\begin{align}\label{6pions}
    A_{6,\text{min}}^{\text{NLSM}} = \frac{1}{2}\frac{(X_{13}+X_{24})(X_{46}+X_{15})}{X_{14}}-X_{13} + \text{cyc.}
\end{align}
Replacing $X_{ij} \mapsto \tau_{ij}$ in the above function exactly matches the leading $\mathcal{O}(\tau_{ij})$ contribution in \eqref{abel6pt}. The higher-order contributions $\sim \tau_{13}\tau_{35}\tau_{15} + \text{cyc.}$ in \eqref{abel6pt} then arise purely from abelianizing the "snowflake" topology corresponding to the second to last diagram in \eqref{6ptDiag}.

\subparagraph{Diagram-level abelianization: $n$ points}
Let us now study the abelianization of a general $n$-point diagram topology. We will label the topology by a list of its factorization channels $x=\lbrace x_{ij}\rbrace$. It will be useful to split the kinematic invariants $\lbrace x_{ij}\rbrace$ into two subsets $x_\text{even}$ and $x_\text{odd}$ corresponding to even- or odd-particle multiplicity channels such that $x=x_\text{even}\cup x_\text{odd}$.
The abelianization of the diagram $x$ according to \eqref{defAbel} is then given by a sum over terms of the form (cf. \eqref{abel61} where $x_{\text{even}}{\,=\,}\lbrace x_{14} \rbrace$ and $x_{\text{odd}}{\,=\,}\lbrace x_{13},x_{15} \rbrace$)
\begin{align}\label{abelGenTop}
 \hspace{-0.1cm}
        \begin{matrix}
        \hspace{-0.1cm}\vspace{-0.1cm}\includegraphics[width=3.7cm]{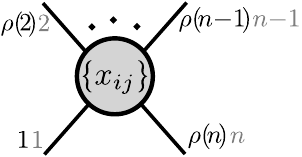}
        \hspace{-0.7cm}\end{matrix} = \frac{\theta(\rho)}{\prod\! x_\text{even}(\rho) \,\prod\! x_\text{odd}(\rho)},
\end{align}
where $x_{ij}(\rho)$ and $\theta(\rho)$ are defined as before. We then observe that the result of abelianizing \eqref{abelGenTop} always takes the simple form
\begin{align}\label{abelGenTopRes}
    \hspace{-0.20cm}\prod\!\!\left(\!\frac{1}{t_{\text{even}}} {\,-\,} \frac{1}{s_{\text{even}}} \!\right)\!\prod\!\!\left(\!\frac{1}{t_{\text{odd}}} {\,+\,} \frac{1}{s_{\text{odd}}} \!\right) = \frac{\prod (-\tau_\text{even})}{\prod \tau_\text{odd}}\,.
\end{align}
Notably, while each term in \eqref{abelGenTop} only involves variables $t_{ij},s_{ij}$, their sum always arranges itself into a function purely of $\tau_{ij}$ by virtue of the trigonometric identities \eqref{trigID1}. The factorized form \eqref{abelGenTopRes} can be proven from the intersection number \cite{Mizera:2017cqs} interpretation of inverse KLT matrix elements $m_n^{\alpha'}$. However, a detailed discussion of the proof goes beyond the scope of this letter.

In \eqref{abelGenTopRes} we see that the abelianization of any diagram topology follows a simple pattern. Kinematic variables $x_\text{even}$ corresponding to even-particle channels are moved to the numerator, while odd-particle channels $x_\text{odd}$ remain in the denominator.

Thus, starting from any diagram topology for the diagonal inverse KLT kernel $m_n^{\alpha'}\!\left[\mathds{1}|\mathds{1}\right]$ (given entirely in terms of $t_{ij}$) we can immediately obtain its abelianization by the formal replacement
\begin{align} \label{abelRep}
    \frac{1}{t_\text{even}}\mapsto -\tau_\text{even}, \hspace{1.0cm} \frac{1}{t_\text{odd}}\mapsto \frac{1}{\tau_\text{odd}}.
\end{align}
We can also apply the replacement \eqref{abelRep} to the diagonal matrix element $m_n^{\alpha'}\!\left[\mathds{1}|\mathds{1}\right]$ as a whole and directly extract the full abelianized function \eqref{defAbel}. Thus, the diagonal matrix element by itself encodes all the information required to describe pion scattering. To make this connection more concrete, we will now show that the formal replacement \eqref{abelRep} can be implemented via a simple kinematic deformation, the $\alpha'$-shift.

\subparagraph{IV. Pions via $\alpha'$-Shift}

We have seen that abelianization naturally distinguishes even- and odd-particle multiplicity channels $x_{\text{even}}$ and $x_{\text{odd}}$ as exemplified by the structure of \eqref{abelGenTopRes} and the replacement rules \eqref{abelRep}. The latter can be realized through a direct manipulation of the stringy kinematic variables \eqref{strPlMand}. Recalling the basic trigonometric identity
\begin{align}\label{trigID2}
    \frac{1}{\tan\left(x \pm \frac{\pi}{2}\right)} = -\tan\left(x \right),
\end{align}
the replacement \eqref{abelRep} can be implemented in a two-step process.
For $n=2k$ even, we consider the diagonal inverse KLT matrix element $m_{2k}^{\alpha'}\!\left[\mathds{1}|\mathds{1}\right]$. First, we rescale 
\begin{align}\label{alphaRes}
   \alpha' \mapsto \alpha'/2, 
\end{align}
which effectively takes $t_{ij}\mapsto \tau_{ij}$ for any $ij$. Then we shift the kinematic $X$-variables according to
\begin{align}\label{Xshift}
    \hat{X}_{\text{even}}\equiv X_{\text{even}}\pm1/\alpha', \hspace{0.5cm} \hat{X}_{\text{odd}} \equiv X_{\text{odd}},
\end{align}
which, via \eqref{trigID2}, effects the replacement \eqref{abelRep}. We will refer to \eqref{Xshift} and the corresponding rescaling of $\alpha'$ as the $\alpha'$-shift. 
Considering now the diagonal inverse KLT kernel $m_{2k}^{\alpha'}\!\left[\mathds{1}|\mathds{1}\right]$ and the abelianized functions $A_{2k}^{\alpha'}$ directly as functions of $X_{ij}$ and $\alpha'$ we have
\begin{align}\label{scalarPionEQ}
\boxed{
    A_{2k}^{\alpha'}(X_{ij}) = m_{2k}^{\alpha'\!\!/2}\!\left[\mathds{1}|\mathds{1}\right](\hat{X}_{ij}).}
\end{align}
This can be read as a statement of \textit{equivalence between cubic scalars and pions}. Indeed, \eqref{scalarPionEQ} asserts that scalars and pions, including a suitable stringy extension, both arise from \textit{the same function} $m_{2k}^{\alpha'}\!\left[\mathds{1}|\mathds{1}\right]$ evaluated on different kinematics and values of $\alpha'$ as in \eqref{alphaRes} and \eqref{Xshift}.

We emphasize that \eqref{scalarPionEQ} is a true equivalence as the $\alpha'$-shift \eqref{Xshift} is invertible. That is, we can start with the abelianized function $A_n^{\alpha'}$, rescale $\alpha' \mapsto 2\alpha'$ and shift $X_{\text{even}}\mapsto X_{\text{even}}\pm1/(2\alpha')$, $X_{\text{odd}}\mapsto X_{\text{odd}}$ to arrive back at the original form of the inverse KLT kernel $m_n^{\alpha'}\!\left[\mathds{1}|\mathds{1}\right]$.

The $\alpha'$-shift \eqref{Xshift} for the kinematic $X$-variables formally closely resembles the $\delta$-shift proposed in \cite{Arkani-Hamed:2024nhp,Arkani-Hamed:2024yvu}. For field theory amplitudes in the $\text{tr}\phi^3$ theory, the $\delta$-shift takes $X_{oo}\mapsto X_{oo}+\delta$, $X_{ee}\mapsto X_{ee}-\delta$ which allows to directly extract pion amplitudes by expanding the shifted scalar amplitudes at $\delta=\infty$. In this context $X_{ee},X_{oo}$ refer to kinematic variables $X_{ij}$ whose labels $ij$ are both even ($ee$) or both odd ($oo$), corresponding exactly to the even-multiplicity channels $X_{\text{even}}$ also shifted by \eqref{Xshift}.

However, we should emphasize that while the $\alpha'$-shift bears a close formal resemblance to the $\delta$-shift, the two are essentially different in the way in which they connect scalars and pions. Nevertheless, the question arises whether both shifts can be seen as vestiges of an underlying structure yet to be discovered in these amplitudes.

Finally, let us remark on a peculiar feature of the $\alpha'$-shift \eqref{Xshift}. Because the diagonal inverse KLT matrix element is purely a function of tangents $t_{ij}$, the trigonometric identity \eqref{trigID2} ensures that the sign in the shift $X_{oo/ee}\mapsto X_{oo/ee}\pm 1/\alpha'$ can be chosen arbitrarily, leaving the result \eqref{scalarPionEQ} unchanged.
This property is not shared by the $\delta$-shift, where the relative signs of the shift for different kinematic variables matter. However, the sign ambiguity of the $\alpha'$-shift will be useful when we want to construct mixed amplitudes that involve both cubic scalars and pions.

\subparagraph{V. Mixed Amplitudes}

We now turn to discuss mixed amplitudes of cubic scalars $\phi$ and pions $\pi$ in a theory commonly referred to as NLSM+$\phi^3$. These amplitudes were first discovered by studying soft limits of pion amplitudes \cite{Cachazo:2016njl} and more recently appear in factorizations of pion amplitudes near so-called ``hidden'' zeros \cite{Arkani-Hamed:2023swr}. Furthermore, various classes of mixed field theory amplitudes can be obtained from an appropriate $\delta$-shift of the tr$\phi^3$ theory \cite{Arkani-Hamed:2024nhp}.
A simple example is the five-point mixed amplitude involving three adjacent scalars $\phi$ and two pions $\pi$,
\begin{align}\label{5ptMixed}
    M_5(\phi\pi\pi\phi\phi) = -\frac{X_{13}+X_{24}}{X_{14}}-\frac{X_{24}+X_{35}}{X_{25}}+1,
\end{align}
which appears in the soft limit and hidden factorizations of the six-point pion amplitude \eqref{6pions}.

Returning to the stringy inverse KLT kernel, we will define stringy mixed functions via a procedure called \textit{semi-abelianization} in analogy to semi-abelianized $Z$-functions \cite{Carrasco:2016ygv}. For us, a general $n$-point mixed function $M_n^{n_\phi,\alpha'}$ will involve $n_\phi$ cubic scalars with labels $\lbrace1\rbrace \cup \phi=\lbrace 1\,\phi_2\dots\phi_{n_\phi}\rbrace$ and $n_\pi=n-n_\phi$ pions labeled $\pi=\lbrace \pi_{1}\dots\pi_{n_\pi} \rbrace$ such that $\phi \cup \pi = \lbrace 2\dots n \rbrace$.
It is defined via semi-abelianization of the inverse stringy KLT kernel,
\begin{align}\label{defSemiAbel}
    M_n^{n_\phi,\alpha'}\!\!\left(1\,\phi_2\dots\phi_{n_\phi}\pi_1\dots \pi_{n_\pi} \right) = \sum\nolimits_{\hat{\rho}}\,m_n^{\alpha'}\!\left[\mathds{1}|1\hat{\rho}\,\right],
\end{align}
where the sum is over $\hat{\rho}=\phi\shuffle\lbrace \pi_{n_1}\rbrace \shuffle \dots \shuffle \lbrace\pi_{n_\pi}\rbrace$ and $\shuffle$ is the shuffle product \cite{Carrasco:2016ygv}. Generically the orderings $\hat{\rho}\subseteq \rho(2\dots n)$ summed over in the semi-abelianization are a subset of those in the full abelianization \eqref{defAbel}. For the special case $\phi=\lbrace\rbrace$ the sums in \eqref{defSemiAbel} and \eqref{defAbel} agree. At low energies, the semi-abelianized functions yield the corresponding field theory mixed amplitudes $M_n^{n_\phi}(1\phi\dots\pi\dots)$,
\begin{align}\label{mixedLOalpha}
    M_n^{n_\phi,\alpha'} \!\!= (\pi\alpha')^k M_n^{n_\phi} + \mathcal{O}(\alpha'^{k+1})
\end{align}
where the leading power in $\alpha'$ is $k=3-n_\phi$ if $n_\pi$ is even and $k=4-n_\phi$ if $n_\pi$ is odd.

In the following, we will show that certain mixed functions \eqref{defSemiAbel} can alternatively be obtained, like the pure pion functions \eqref{scalarPionEQ}, from various $\alpha'$-shifts of the diagonal inverse KLT matrix element $m_n^{\alpha'}\!\left[\mathds{1}|\mathds{1}\right]$. 
Since the space of all mixed functions is vast, we cannot be comprehensive and instead give illustrative examples.
Throughout we will only study mixed functions involving an even number of pions $n_\pi$ as they correspond to stringy extensions of the NLSM+$\phi^3$ amplitudes discussed originally in \cite{Cachazo:2016njl}.

To arrive at suitable $\alpha'$-shifts for our subsequent discussion of mixed functions, we follow the prescription of \cite{Arkani-Hamed:2024nhp} for the $\delta$-shift regarding which $X$-variables need to be shifted to obtain a given mixed amplitude. For our purposes we then replace $\delta\to \pm 1/\alpha'$ where we are free to choose signs thanks to \eqref{trigID2}. In addition, we always apply a rescaling $\alpha' \mapsto \alpha'/2$ as in the case of \eqref{alphaRes}.

\subparagraph{Three Scalars}

The first set of stringy mixed functions we study are those with $n=2k+1$ odd involving three adjacent scalars $M_{2k+1}^{3,\alpha'}(\phi\pi\dots\pi\phi\phi)$. These can be obtained by applying the known shift \eqref{Xshift} for pions to the odd-point inverse KLT kernel, i.e. in analogy to \eqref{scalarPionEQ} we now have
\begin{align}\label{scalar3phiEQ}
    M_{2k+1}^{3, \alpha'}(X_{ij}) = m_{2k+1}^{\alpha'\!\!/2}\!\left[\mathds{1}|\mathds{1}\right](\hat{X}_{ij}).
\end{align}
To give an explicit example, we apply the shift \eqref{Xshift} to the five-point diagonal matrix element in \eqref{invKLTex} and obtain
\begin{align}
    M_{5}^{3, \alpha'} \!= -\frac{\tau_{13}+\tau_{24}}{\tau_{14}}-\frac{\tau_{24}+\tau_{35}}{\tau_{25}}+1+\tau_{13}\tau_{35}.
\end{align}
At leading order in $\alpha'$ this can be seen to yield the known mixed amplitude \eqref{5ptMixed}. More generally, at $2k{+}1$ points the mixed functions \eqref{scalar3phiEQ} agree with the stringy functions obtained via semi-abelianization \eqref{defSemiAbel} for $\phi=\lbrace 2k,2k{+}1\rbrace$ and $\pi=\lbrace 2\dots 2k{-}1\rbrace$.

As our next example we discuss the nine-point function $M_{9}^{\alpha'}\!(\phi\pi\pi\phi\pi\pi\phi\pi\pi)$. In \cite{Arkani-Hamed:2024nhp} it was shown that this particular mixed amplitude cannot be obtained from a $\delta$-shift as there is no consistent choice of signs $X_{ij}\pm\delta$ for the $X$-variables involved. However, since the $\alpha'$-shift is agnostic about these signs due to \eqref{trigID2} we can compute this amplitude from the diagonal matrix element $m_9^{\alpha'}\!\left[\mathds{1}|\mathds{1}\right]$ in the stringy setting. We rescale $\alpha'\to \alpha'/2$ and shift $\tilde{X}=\lbrace X_{13}+\text{cyc.},X_{16},X_{49},X_{37},X_{26},X_{59},X_{38}\rbrace$. Here the set notation $\lbrace X_{ij}\rbrace$ indicates that we shift the listed variables according to $\tilde{X}_{ij} = X_{ij}\pm 1/\alpha'$ and leave all other $X$-variables unchanged. Under this shift we obtain
\begin{align}\label{scalar3phi9ptEQ}
     M_{9}^{ \alpha'}\!\!(\phi\pi\pi\phi\pi\pi\phi\pi\pi)(X_{ij}) = m_{9}^{\alpha'\!\!/2}\!\left[\mathds{1}|\mathds{1}\right](\tilde{X}_{ij}),
\end{align}
which yields the correct field theory amplitude in the limit $\alpha'\to 0$ and agrees with the semi-abelianized function \eqref{defSemiAbel} setting $\phi=\lbrace4,7 \rbrace$ and $\pi=\lbrace 2,3,5,6,8,9\rbrace$.

More generally we find that up to nine points stringy functions with three scalars can be equivalently obtained from an $\alpha'$-shift or semi-abelianization.

\subparagraph{Four and more Scalars}
While the $\alpha'$-shift prescription and the semi-abelianization \eqref{defSemiAbel} agree for stringy functions with exactly three scalars, the story is subtler once we study stringy functions with four or more scalars.

To illustrate this we consider the particular example of $M_{6}^{\alpha'}\!(\phi\phi\phi\phi\pi\pi)$. From \cite{Arkani-Hamed:2024nhp} we gather that a suitable shift for this mixed function is taking $\tilde{X}_{ij}\pm 1/\alpha'$ with $\tilde{X}=\lbrace X_{15},X_{26},X_{36},X_{46}\rbrace$. Applying the shift to the inverse KLT kernel we obtain $M_{6,\text{shift}}^{\alpha'}(X_{ij}) = m_6^{\alpha'\!\!/2}\!\left[\mathds{1}|\mathds{1}\right](\tilde{X}_{ij})$.
Alternatively we can compute the semi-abelianized function $M_{6,\text{abel}}^{\alpha'}$ via \eqref{defSemiAbel} where $\phi=\lbrace 2,3,4\rbrace$ and $\pi=\lbrace 5,6\rbrace$.

Now at the leading order $\mathcal{O}(1/\alpha')$ the functions $M_{6,\text{shift}}^{\alpha'}$ and $M_{6,\text{abel}}^{\alpha'}$  agree (up to normalization) and yield the expected field theory mixed amplitude \cite{Carrasco:2016ygv},
\begin{align}\label{M6LO}
    \lim_{\alpha'\to 0} M_{6,\text{abel}}^{\alpha'} = \frac{1}{2}\lim_{\alpha'\to 0} M_{6,\text{shift}}^{\alpha'} = \frac{1}{\pi\alpha'}\, M_6(\phi\phi\phi\phi\pi\pi).
\end{align}
At finite $\alpha'$ both functions can be expressed entirely in terms of the stringy variables $\tau_{ij}$. Comparing them directly we find
\begin{align}\label{M6AbelShiftDelta}
    M_{6,\text{abel}}^{\alpha'}(\tau_{ij}) = \frac{1}{2}M_{6,\text{shift}}^{\alpha'}(\tau_{ij}) + \mathcal{O}(\tau_{ij}),
\end{align}
i.e. the stringy function $M_{6,\text{shift}}^{\alpha'}$ obtained via $\alpha'$-shift is missing certain higher-order corrections relative to $M_{6,\text{abel}}^{\alpha'}$, which, however, have no bearing on the low-energy limit \eqref{M6LO}.

This mismatch between mixed stringy functions $M_{n,\text{abel}}^{\alpha'}$ and $M_{n,\text{shift}}^{\alpha'}$ obtained from semi-abelianization and the $\alpha'$-shift persists more generally. However, we have verified for mixed functions up to ten points with six scalars $\phi$ that both approaches yield the same mixed amplitudes at leading order in $\alpha'$.
%

\subparagraph{VI. Discussion and Future Directions}
The interchangeability of pions and cubic scalars through the $\alpha'$-shift \eqref{Xshift} of the inverse string theory KLT kernel $m_n^{\alpha'}$ advocated here opens up a number of avenues for further study.

Let us start by highlighting the monodromy properties \cite{Mizera:2016jhj} of the stringy inverse KLT kernel. The matrix elements $m_n^{\alpha'}$ satisfy monodromy relations
\begin{align}\label{invKLTmono}
    m_n^{\alpha'}\!\left[\mathds{1}|\mathds{1}\right] + \sum_{k=2}^{n-1}e^{ix_k} m_n^{\alpha'}\!\left[2\dots k\,1\,k{+}1\dots n|\mathds{1}\right] = 0,
\end{align}
with phases given by $x_k=\pi\alpha'2p_1\!\cdot\!(p_2+\dots+p_k)$. For $\alpha'\to 0$ the relations \eqref{invKLTmono} are known to reduce to the Kleiss-Kuijf (KK) and Bern-Carrasco-Johannson (BCJ) relations for amplitudes of the BAS theory.

An immediate implication of \eqref{invKLTmono} for the abelianized functions $A_n^{\alpha'}$ is that they satisfy monodromy too, as they correspond to linear combinations of inverse KLT matrix elements \eqref{defAbel}. Consequently, both stringy scalars and pions exhibit certain ``hidden'' zeros \cite{Arkani-Hamed:2023swr,Li:2024qfp,Bartsch:2024amu} as well as the larger class of monodromy zeros \cite{DAdda:1971wcy}. 
The inverse KLT kernel and abelianized pion functions described here provide simple, rational functions suitable for explicitly studying these zeros. For instance, it would be interesting to see if there is a factorization behavior of these functions near the monodromy zeros similar to the one observed for the ``hidden'' zeros in field theory amplitudes \cite{Arkani-Hamed:2023swr,Cao:2024gln}. 

A further possible connection is to positive geometries. The field theory limit of the stringy KLT kernel corresponds to amplitudes in the BAS theory, which are described by the ABHY associahedron \cite{Arkani-Hamed:2017mur}. It is tempting to ask whether their geometry can be generalized to the $\alpha'$-complete inverse string theory KLT kernel $m_n^{\alpha'}$, and whether, via the $\alpha'$-shift presented here, it can be extended to pions and mixed amplitudes. The recent discovery of such structures for special classes of mixed field theory amplitudes \cite{Paranjape:2025wjk} suggests an affirmative answer.

Another direction is to explore soft/KLT/zeroes bootstrap methods \cite{Cheung:2014dqa,Cheung:2016drk,Elvang:2018dco,Kampf:2019mcd,Chi:2021mio,Cheung:2021yog,Brown:2023srz,Kampf:2024xjy,Bartsch:2024ofb} in this context, incorporating the $\alpha'$-dependence in the kinematical constraints.

Finally, ongoing work by the authors suggests that the close connection between the inverse string theory KLT kernel and pions also extends to loop integrands.

To summarize, the inverse KLT kernel $m_n^{\alpha'}$ and the stringy pions $A_n^{\alpha'}$ exhibit many properties that have received great attention in the recent literature \cite{Arkani-Hamed:2023swr,Arkani-Hamed:2024yvu,Arkani-Hamed:2024nhp,Arkani-Hamed:2024vna,Arkani-Hamed:2024pzc,Arkani-Hamed:2023jry,Arkani-Hamed:2024tzl,Li:2024qfp,Bartsch:2024amu,Cao:2024gln,Arkani-Hamed:2017mur,Paranjape:2025wjk,Li:2024bwq,Arkani-Hamed:2024fyd,Rodina:2024yfc,Backus:2025hpn}, all while being simple rational functions of stringy kinematic variables. Therefore, they promise to provide a valuable testing ground to probe for novel features of amplitudes that go beyond those discussed above. This in turn could meaningfully inform the study of other string-related amplitudes, for instance those involving gluons or gravity.

\medskip

{\it Acknowledgements}: We thank Nima Arkani-Hamed, Song He, Shruti Paranjape, and Jonah Stalknecht for useful discussions. This work is supported by GA\-\v{C}R 24-11722S, MEYS LUAUS23126, OP JAK CZ.02.01.01/00/22\_008/0004632, DOE grant No. SC0009999 and the funds of the University of California.
\vspace{-0.55cm}
\bibliography{mainbib}
\bibliographystyle{apsrev4-1}

\end{document}

%% file: header_data.tex
\titleformat{\section}{\centering\normalsize\normalfont\bf}{\thesection}{0em}{}
\hypersetup{pdftitle={},pdfcreator={},linkcolor=[rgb]{0.15,0.35,0.75},colorlinks=true,citecolor=[rgb]{0.675,0,0.2},urlcolor=[rgb]{0.15,0.35,0.65}}
\renewcommand{\hat}{\widehat}
\renewcommand{\tilde}{\widetilde}

\definecolor{nmhv}{rgb}{0.95,0.55,0.55}
\definecolor{mhvblue}{rgb}{0.6,0.6,0.7765}
\definecolor{ampgrey}{rgb}{0.9,0.9,0.9}
\definecolor{hblue}{rgb}{0,0,0.575}
\definecolor{hred}{rgb}{0.575,0.0,0.225}
\definecolor{hgreen}{rgb}{0.0,0.4,0.2}
\definecolor{hteal}{rgb}{0.0,0.545,0.7451}
\definecolor{perm}{rgb}{0.1,0.45,0.85}
\definecolor{unord}{rgb}{0,0,0}
\definecolor{ord}{rgb}{0,0,0.575}
\definecolor{anchorLeg}{rgb}{0.575,0.0,0.225}
\definecolor{fRed}{rgb}{0.48,0.02824,0.18824}